\documentclass[aps,pra,twocolumn,showpacs,floatfix]{revtex4}

\usepackage{amsmath} % AMS mathematical facilities for LaTeX.
\usepackage{amssymb} % AMS symbol fonts for LaTeX.
\usepackage{graphicx}

\newcommand{\op}[1]{\hat{\rm{#1}}}

\def\Berlin{
     AG Moderne Optik,
     Institut f\"ur Physik,
     Humboldt-Universit\"at zu Berlin,
     Newtonstr. 15, D\,-\,12\,489 Berlin, Germany
}

\begin{document}

%\preprint{}

\title{Imaging of the umbrella motion and tunneling in the ammonia molecule 
by strong-field ionization}

\author{Johann F\"orster}

\author{Etienne Pl\'{e}siat\footnote{Present address: CIC nanoGUNE, E-20018 Donostia-San Sebasti{\'a}n, Spain.}} 

\author{{\'A}lvaro Maga\~{n}a} 

\author{Alejandro Saenz} 

\affiliation{\Berlin}

\date{\today}

\begin{abstract}
  The geometry-dependent ionization behavior of the ammonia molecule
  is investigated.  Different theoretical approaches for obtaining the
  ionization yield are compared, all of them showing a strong
  dependence of the ionization yield on the inversion coordinate at
  long wavelengths ($\geq 800$ nm). It is shown how this effect can be
  exploited to create and probe nuclear wave packets in neutral
  ammonia using \textit{Lochfra\ss}.  Furthermore, imaging of a wave packet
  tunneling through the barrier of a double-well potential in
  real time is discussed.
\end{abstract}

% \keywords{}

\maketitle

%
%#############################################################################
\section{Introduction}
%#############################################################################
%

The intense, ultrashort laser pulses that became accessible during the
last decade offer prospects to manipulate and image molecules on their
natural scales (sub-femtosecond and sub-\aa ngstr\"om).  Investigating
the response of small molecules to these laser fields, especially
high-harmonic generation HHG and ionization (seen as the first step of
HHG), led to the development of promising concepts to produce a
real-time movie of the electronic and nuclear dynamics triggered in
these molecules.  The information contained in high-harmonic radiation
emitted from these molecules may be used for, e.g., orbital tomography
\cite{sfm:itat04}, probing of nuclear dynamics with sub-fs resolution
\cite{sfm:lein05,sfm:bake06,sfm:farr11a,sfm:krau13,sfm:foer13}, and
imaging coupled electron-nuclear dynamics \cite{sfm:worn10c,
  sfm:woer11}.  Furthermore, also the electrons emitted by ionization 
contain structural information suitable for orbital imaging
\cite{sfm:meck08, sfm:petr10a}.  Since nuclear motion may strongly
influence the ionization behavior, these electrons also contain
information about the nuclear dynamics as discussed in the following.

For the example of molecular hydrogen it was found experimentally that
the transition from the neutral molecule to the molecular ion may not
follow the Franck-Condon distribution \cite{sfm:urba04}.  This
experiment confirmed an earlier prediction \cite{sfm:saen00c} in which
the effect was explained assuming a much faster response of the
electrons to a laser field compared to the response of the nuclei.
Consequently, a vertical electronic transition occurs at fixed nuclear
geometries.  The breakdown of the Franck-Condon approximation stems
from the exponential dependence of the strong-field ionization rates
on the binding energy of the ejected electron \cite{sfa:pere66,
  sfa:ammo86, sfa:ilko92} (ionization potential of the initial state).
Already small changes in the binding energy lead to a very pronounced
dependence of the electronic transition amplitude on nuclear geometry
\cite{sfm:saen00c}. This is the case when the equilibrium geometry of
the initial state of the neutral molecule and the final state of the
molecular ion are significantly different.  This prediction of a possibly 
strong internuclear-separation dependent ionization probability was later 
further confirmed by explicit solutions of the time-dependent
Schr\"odinger equation of H$_2$ \cite{sfm:awas06, sfm:vann09,
  sfm:foer14}.  A process termed \textit{Lochfra\ss} \cite{sfm:goll06}
was theoretically proposed as a method which exploits the extreme
nuclear-geometry dependence of strong-field ionization rates to create
and measure nuclear wave packets in the electronic initial states of
neutral molecules. Indeed, \textit{Lochfra\ss} was found to be the
dominant mechanism (compared to bond softening) in experiments 
imaging a vibrational wave packet in
neutral diatomic molecules with sub-femtosecond and sub-\aa ngstr\"om
resolution \cite{sfm:ergl06, sfm:fang08a, sfm:fang08b}.  Remarkably,
it was possible to follow the nuclear wave packet experimentally for
an extremely long time window (larger than 1 ps) \cite{sfm:ergl06}, 
leading to the proposal of a molecular clock.  Furthermore, this mechanism was
shown to orient polar molecules due to the rotational wave packet
created by orientation-dependent ionization \cite{sfm:span12, sfm:frum12b}.

The ammonia molecule undergoes a significant equilibrium geometry
change (pyramidal to plane) upon ionization and is thus a good
candidate for observing Lochfra\ss\ in a molecule beyond diatomics.
Recently, the relative nuclear motion launched in the cations NH$_3^+$
and ND$_3^+$ was studied and experimentally observed by comparison of
the respective high-harmonic spectra \cite{sfm:foer13,sfm:krau13}
using the so-called PACER (Probing Attosecond dynamics with
Chirp-Encoded Recollisions) technique \cite{sfm:lein05, sfm:bake06}.
It was shown that the required nuclear autocorrelation functions can
be obtained from photoelectron spectra \cite{sfm:farr11a,
  sfm:foer13,sfm:krau13}. The photoelectron spectra revealed that the
inversion (or umbrella) mode is the only vibrational mode in which a
significant nuclear motion is triggered by the laser field
\cite{sfm:krau13}.  Importantly, it was shown that the strong
geometry-dependence of the ionization rate, the key ingredient for
Lochfra\ss, must indeed be considered in order to quantitatively match
the experimental PACER results \cite{sfm:krau13, sfm:foer13}.

In the following section we briefly discuss the methods and basis-set
parameters used to describe the electronic response (ionization) and
triggered nuclear motion in neutral ammonia.  In
Sec.~\ref{sec:ion-and-lochfrass} we discuss the observed ionization
behavior and its application to imaging based on Lochfra\ss.
Sec.~\ref{sec:tunneling-wavepacket} introduces the idea to image a
tunneling wave packet.  This idea is further fleshed out in
Sec.~\ref{sec:asymmetric-ionization} where the observation of the
tunneling process in real time and the required laser fields /
ionization behavior to achieve this goal are discussed.

%
%#############################################################################
\section{Methods}
%#############################################################################
%

In order to study nuclear motion in neutral ammonia, we first describe
the geometry-dependent electronic response (ionization) of the
molecule to the laser field. Thereafter, this response is included in the
description of the (slower) nuclear motion. Unless noted otherwise,
atomic units with $\hbar=e=m_e=4\pi \epsilon_0 = 1$ are adopted.

The time-dependent Schr\"odinger equation (TDSE) describing the
electronic response for a fixed nuclear geometry is solved using the
single-determinant approach which has been described in detail in our
previous works \cite{sfm:awas08, sfm:petr10a,sfm:farr11a,
  sfm:petr13a}.  Briefly, a multi-center B-spline approach is used to
solve the Kohn-Sham hamiltonian. The ground-state density calculated
from a DFT calculation employing the ADF quantum chemistry software
\cite{gen:adf} is employed to construct the Kohn-Sham hamiltonian
which is then fully diagonalized in a multicentric B-spline basis.  We
then time-propagate in the basis of these orbitals without freezing
any of the lower-lying states in the expansion.  In the case
considered here in which ionization from the 3$a_1$ HOMO orbital
dominates the ionization yield, the total yield is calculated as the
sum over the individual populations $P_j = \left|c_j\right|^2$ of the
continuum orbitals (with energy $E \geq E_{\mathrm{thr}}$) after the
laser field,
\begin{equation}
  Y \quad = \sum_{j;\; E_j \geq E_{\mathrm{thr}}} P_j \quad . \label{eq:ionization}
\end{equation}

\begin{figure}
\begin{center}
\includegraphics[width=0.18\textwidth]{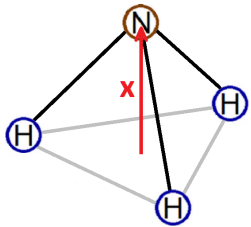}\\
\caption{\label{fig:nh3_structure} (Color online) Geometry of the
  NH$_3$ molecule. The inversion coordinate $x$ (red arrow) describes
  the distance of the nitrogen atom to the center of the triangle
  spanned by the three hydrogen atoms (for a negative $x$ the nitrogen
  atom is located below the triangle).}
\end{center}
\end{figure}
In order to describe the inversion (or umbrella) motion of ammonia,
the main nuclear coordinate considered here is the inversion
coordinate $x$, see Fig.~\ref{fig:nh3_structure}. We calculate the
electronic basis for different values of this inversion coordinate
using exactly the same nuclear geometries as in table 1 of
Ref.~\cite{gen:aqui98}.  A box size of $300$ a.u.\ and a central
expansion of $449$ B-spline functions (times a spherical harmonic) is
used. The maximum angular momentum included is $l_{\mathrm{max}}=11$.
In order to check the convergence of the results, we performed a
reference calculation with a larger box of size $600$ a.u., $776$
$B$-spline functions and maximum angular momentum $l=8$. With this
basis, we checked the ionization yields for three different
geometries, namely $x=0$, $0.592$, and $1.066$ a.u.\ and found no
deviations from the results obtained with the smaller box size on the
level of the graphs shown in the following.

The linearly polarized laser field is described classically by the
vector potential
\begin{eqnarray}
  \vec{A}(t) = \left\{\begin{array}{cl} \vec{A}_0 \cos^2(\pi t/T)
      \sin(\omega t + \varphi) & \mbox{for }|t|\leq T/2\\ 0 &
      \mbox{elsewhere} \end{array}\right.
\end{eqnarray}
with laser frequency $\omega=2\pi c/\lambda$ (wavelength $\lambda$,
speed of light $c$), carrier-envelope phase $\varphi$, and total pulse
duration $T=2\pi n_{\mathrm{c}}/\omega$ (number of cycles $
n_{\mathrm{c}}$).  In the following $\varphi=0$ is used.  The
interaction potential reads $\op{V}(t) = \hat{\mathbf{p}} \cdot
\vec{A}(t)$ (dipole approximation, velocity gauge). The polarization
vector $\vec{A}_0$ points along the inversion axis (and thus the
static dipole moment) of ammonia. Ionization from the $3a_1$ HOMO 
orbital is geometrically preferred for this orientation. This 
orientation has been already experimentally realized and variation of 
the orientation did not reveal contributions to the ionization from lower
lying orbitals \cite{sfm:krau13}.

The TDSE ionization yields are compared to the ones obtained using
Ammosov-Delone-Krainov (ADK) tunneling rates $\Gamma_\textrm{ADK}$
\cite{sfa:ammo86, sfa:ilko92} and the recently introduced
frequency-corrected ADK (FC-ADK) model \cite{sfm:foer14} based on the
Perelomov-Popov-Terent'ev (PPT) \cite{sfa:pere66} theory. By
integrating the tunneling rate over the whole pulse duration, one
obtains the ionization yield
\begin{equation}
  Y(x) =
  1-\exp\left\{-\int\!\Gamma[F_{\mathrm{e}}(t),I_p(x)] dt\right\}
\label{eq:Y_ADK}
\end{equation}
%---------------
where $F_{\mathrm{e}}(t)$ is the envelope function of the electric
field and $I_p(x)$ is the vertical binding energy. For consistency, we
use the same geometry-dependent $I_p(x)$ as in the TDSE calculations
(from the diagonalization of the Kohn-Sham hamiltonian in the B-spline
basis).  The FC-ADK ionization rate is defined as \cite{sfm:foer14}
%---------------
\begin{eqnarray}
  \Gamma_{\rm FC-ADK} = N_e \sqrt{\frac{3 F_{\mathrm{e}}}{\pi \kappa^3}
    (2/\kappa -1)} \frac{F_{\mathrm{e}}}{8\pi} \left(\frac{4 e \kappa
      ^ 3}{(2/\kappa -1) F_{\mathrm{e}}}\right)^{2/\kappa}
  \nonumber\\ \times \exp\left[ -\frac{2\kappa^3}{3
      F_{\mathrm{e}}}\,\, g(\gamma) \right] \quad
\label{eq:IonRateCor}
\end{eqnarray}
%---------------
where $e=2.718...\,$, $\kappa = \sqrt{2 I_p(R)}$, $\gamma = \kappa
\omega / F_{\mathrm{e}}$ and $N_e$ is the number of active electrons.
For the H$_2$ molecule it has been shown in \cite{sfm:awas08} that the
single-active electron ionization yield matches the full two-electron
solution when it is multiplied by a factor $2$ ($N_e=2$ active
electrons) if the ionization yield is small $(Y \lesssim
0.1)$. Otherwise, the ionization yield should be left unchanged
($N_e=1$ active electron).  In the following, all ionization yields
obtained within the present TDSE approach and (FC-)ADK for intensities
lower than $I=10^{14}$ W/cm${}^2$
are thus multiplied by a factor $2$.\\
FC-ADK differs from ADK only by the function $g(\gamma)$
\cite{sfa:pere66}
%---------------
\begin{equation}
  g(\gamma) = \frac{3}{2\gamma}\biggl\{
  \Bigl(1+\frac{1}{2\gamma^2}\Bigr)\mathrm{arcsinh}\gamma -
  \frac{\sqrt{1+\gamma^2}}{2\gamma} \biggr\}\,
\label{eq:fun_g}
\end{equation}
(in the limit $\gamma \ll 1$, $g(\gamma)$ approaches $1$ and thus
FC-ADK becomes identical to standard ADK).
%---------------

If a significant part of the initial nuclear wave function gets
ionized, as it is the case for Lochfra\ss, the effect of the
ionization process on the nuclear motion in the electronic ground
state of the neutral molecule is incorporated by means of a loss
channel (ignoring a possible recombination) \cite{sfm:goll06}.  This
leads to the one-dimensional time-dependent Schr\"odinger equation for
the inversion coordinate $x$,
\begin{align}
  i\, \frac{\partial}{\partial t}\, \chi(x,t)\, =\,
  \left(\op{H}_0\, -\, \frac{i\, \Gamma (x,t)}{2}\right)\,
  \chi(x,t)\quad ,\label{eq:lochfrass-prop}
\end{align}
where $\op{H}_0$ is the time-independent nuclear Hamiltonian and
$\Gamma(x,t)$ is the geometry-dependent ionization rate.  In the limit
of an extremely fast ionization process compared to the time scale of
nuclear motion (instantaneous ionization/vertical transition), an
initial nuclear wave function $\chi_0(x)$ is transformed into 
\begin{equation} 
\sqrt{1-Y(x)}\,\chi_0(x) \quad . \label{eq:sudden_approx}
\end{equation} 
In the following, this limit is termed
sudden ionization approximation.  To obtain an ionization rate
$\Gamma_{\mathrm{TDSE}}(x,t)$ for the electronic TDSE calculations, we
assume the time-dependence of FC-ADK and scale the rate with a
constant factor $q(x)$, $\Gamma_{\mathrm{TDSE}}(x,t)\,=\,
q(x)\;\Gamma_{\mathrm{FC-ADK}}(x,t)$. This is done such that the
integration of $\Gamma_{\mathrm{TDSE}}(x,t)$ according to
Eq.~\eqref{eq:Y_ADK} gives the correct ionization yield
$Y_{\mathrm{TDSE}}(x)$.

For the time-independent umbrella motion, we follow the approach in
Ref.~\cite{gen:aqui98} and adopt the position-dependent mass
Hamiltonian \footnote{In contrast to Ref.~\cite{gen:aqui98} we choose
a hermitian ordering for the kinetic energy operator. As has been 
found in chapter IV A of Ref.~\cite{gen:foer12} the eigenenergies 
are practically independent on the choice where $\mu(x)$ enters
the hamiltonian.}
\begin{align}
  \op{H}_0\,=\,\frac{1}{2}\,\op{p}\,\frac{1}{\mu(x)}\,\op{p}\,+\,
  V(x) \label{eq:PDM-hamiltonian}
\end{align}
where $V(x)$ is the double-well potential from table~1 of
Ref.~\cite{gen:aqui98} and
\begin{align}
  \mu(x)\,=\,\frac{3m_{\mathrm{H}}\,m_{\mathrm{N}}}{3m_{\mathrm{H}}\,+\,m_{\mathrm{N}}}\,+\,\frac{3m_{\mathrm{H}}\,x^2}{r_0^{\,2}\,-\,x^2}
\end{align}
is the position-dependent mass where $r_0=1.0041$ a.u.\ and
$m_{\mathrm{H}}$ ($m_{\mathrm{N}}$) denote the mass of the hydrogen
(nitrogen) atom, respectively.  The position-dependent mass
effectively includes the change of the size of the triangle spanned by
the three hydrogen atoms during the umbrella motion (variation of
$x$).  Remarkably, the energies of the eight lowest lying vibrational
states differ by at most $5.6\%$ from their experimental values,
compared to $32.4\%$ if one would adapt a constant mass instead
\cite{gen:aqui98,gen:foer12}.  Equation~\eqref{eq:PDM-hamiltonian} and
its solution have been discussed in detail in \cite{gen:foer12}.
For the solution of Eqs.~\eqref{eq:PDM-hamiltonian} and
\eqref{eq:lochfrass-prop} using the method described in
\cite{gen:foer12} (and its time-dependent formulation using
fast-fourier transformations), we use a box size of
$x_{\mathrm{max}}=\pm 1.805$ a.u.\ and $N=71$ points (converged
results are obtained already with $N=31$ points).

\begin{figure}
\begin{center}
\includegraphics[width=0.49\textwidth]{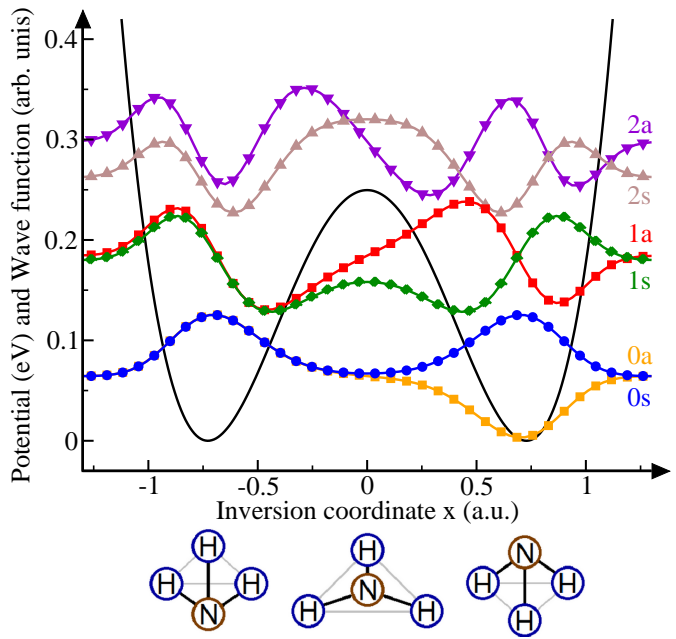}\\
\caption{\label{fig:PotentialAndStates} (Color online) Electronic
  ground-state potential of NH$_3$ (full black line) and vibrational
  eigenfunctions (full colored lines with symbols) labelled by
  symmetry (s/a) with respect to a reflection at $x=0$.  The
  eigenfunctions are vertically shifted such that they approach their
  respective eigenenergies in the limit $|x|\rightarrow\infty$.}
\end{center}
\end{figure}
Fig.~\ref{fig:PotentialAndStates} shows the six lowest-lying
vibrational eigenfunctions of NH$_3$ obtained with the present method.
The energy splitting between the first two vibrational (inversion)
eigenstates is extremely small ($\approx 0.8$ cm${}^{-1}$
\cite{gen:spir83, gen:aqui98, gen:foer12}).  Thus, if the ammonia
molecule is initially in thermal equilibrium, the initial state is (as
also discussed in \cite{sfm:foer13}) a thermal mixture consisting of
$50\%$ in the lowest symmetric state $\left|0s\right\rangle$ and 
$50\%$ in the lowest anti-symmetric state $\left|0a\right\rangle$ for
a large temperature regime (4 K $< T <$ 300 K). This requires a
density matrix formulation for the evolution of the vibrational wave
packet.  The solution simplifies, however, if the inversion symmetry
is not broken at any time, $\Gamma(x,t) = \Gamma(-x,t)$.  In this
case, one can numerically solve the Schr\"odinger
equations~\eqref{eq:lochfrass-prop} independently for
$\left|0s\right\rangle$ or $\left|0a\right\rangle$ as initial state,
respectively, instead of propagating the whole density matrix.

%
%#############################################################################
\section{Ionization behavior and application to imaging
  (Lochfra\ss)}\label{sec:ion-and-lochfrass}
%#############################################################################
%

\begin{figure*}
  \begin{center}
    \includegraphics[width=0.95\textwidth]{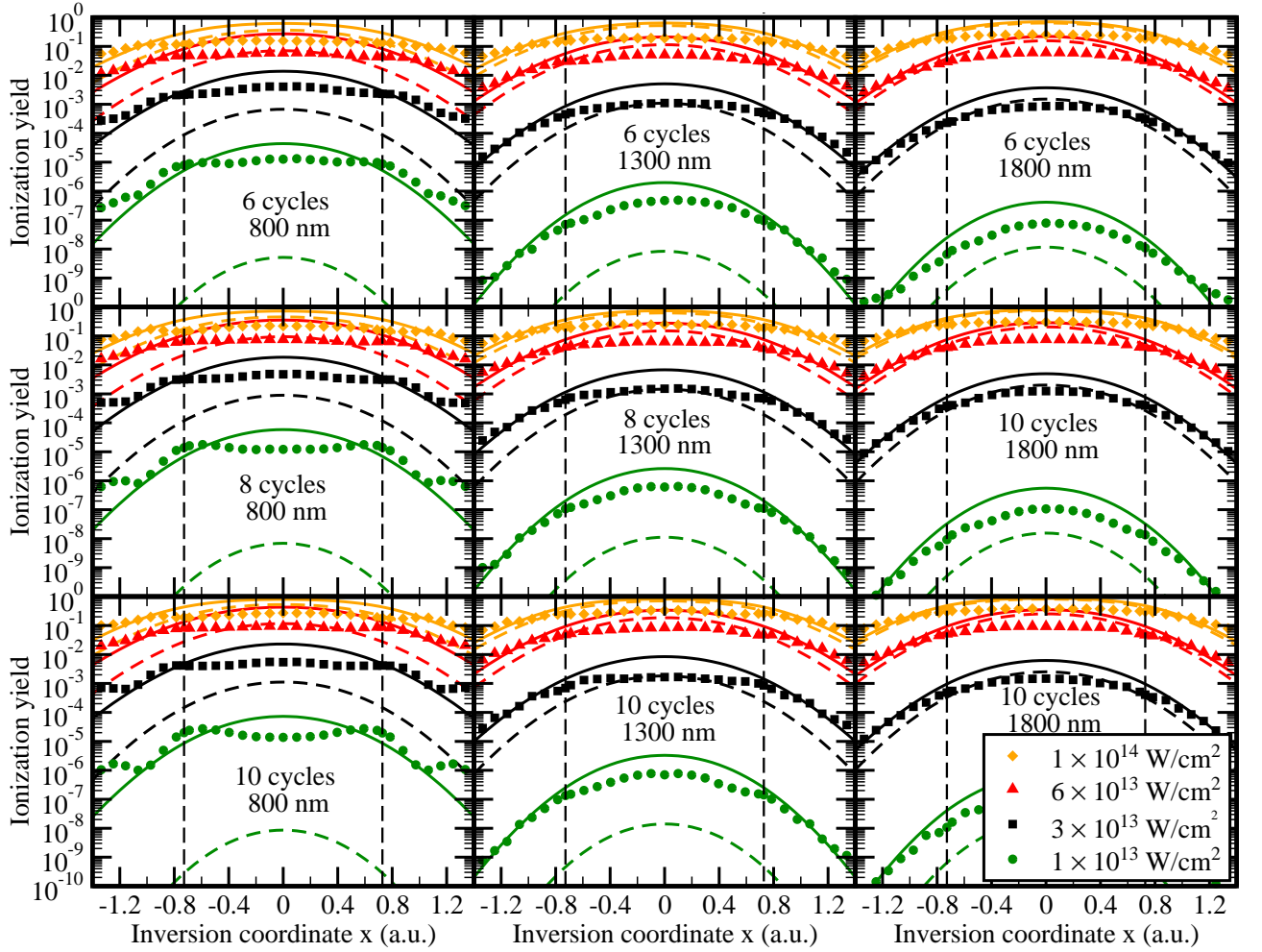}\\
    \caption{\label{fig:IonizationYieldsLong} (Color online)
      Geometry-dependent ionization yields obtained from TDSE
      (points), ADK (dashed lines), and FC-ADK (full lines) for
      different intensities, wavelengths, and number of laser cycles
      (values are given inside the graph). The dashed vertical lines
      indicate the equilibrium geometry at $x_{\mathrm{eq}}=\pm 0.728$
      a.u.}
\end{center}
\end{figure*}

\begin{figure}
\begin{center}
  \includegraphics[width=0.45\textwidth]{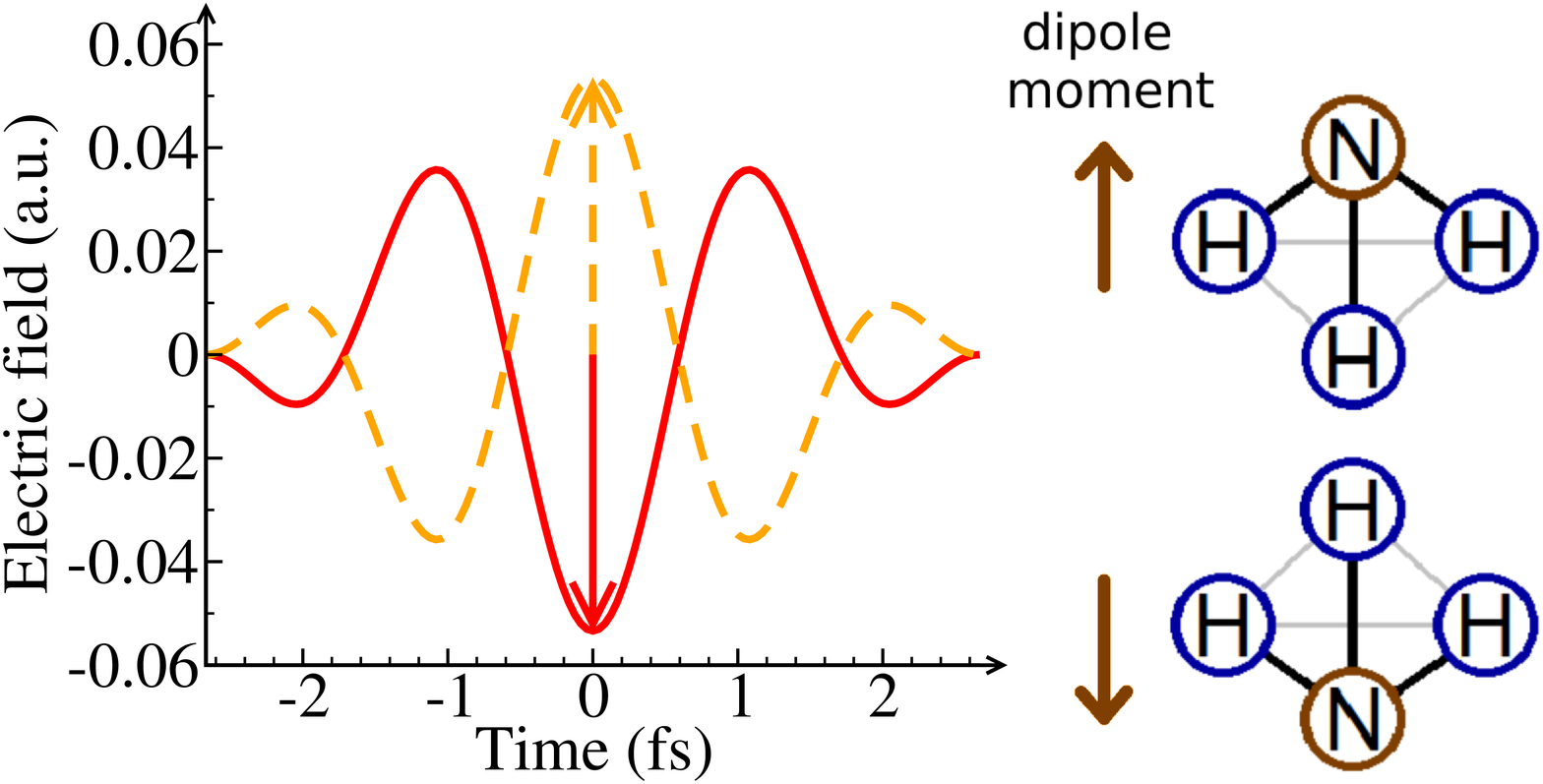}\\
  \caption{\label{fig:dipole_effect} (Color online) The
    time-dependence of an electric field with $n_c=2$ cycles total
    duration, wavelength $\lambda=800$ nm, peak intensity $I=10^{14}$
    W/cm${}^2$ and carrier-envelope phase $\varphi=0$ ($\varphi=\pi$)
    is shown left as dashed orange (full red) line.  For an electric
    field pointing along the inversion axis going through the nitrogen
    atom and the center of the triangle spanned by the hydrogen atoms,
    the maximum of the field may point parallel or antiparallel to the
    dipole moment of the molecule as indicated on the right.}
\end{center}
\end{figure}

\begin{figure}
\begin{center}
\includegraphics[width=0.38\textwidth]{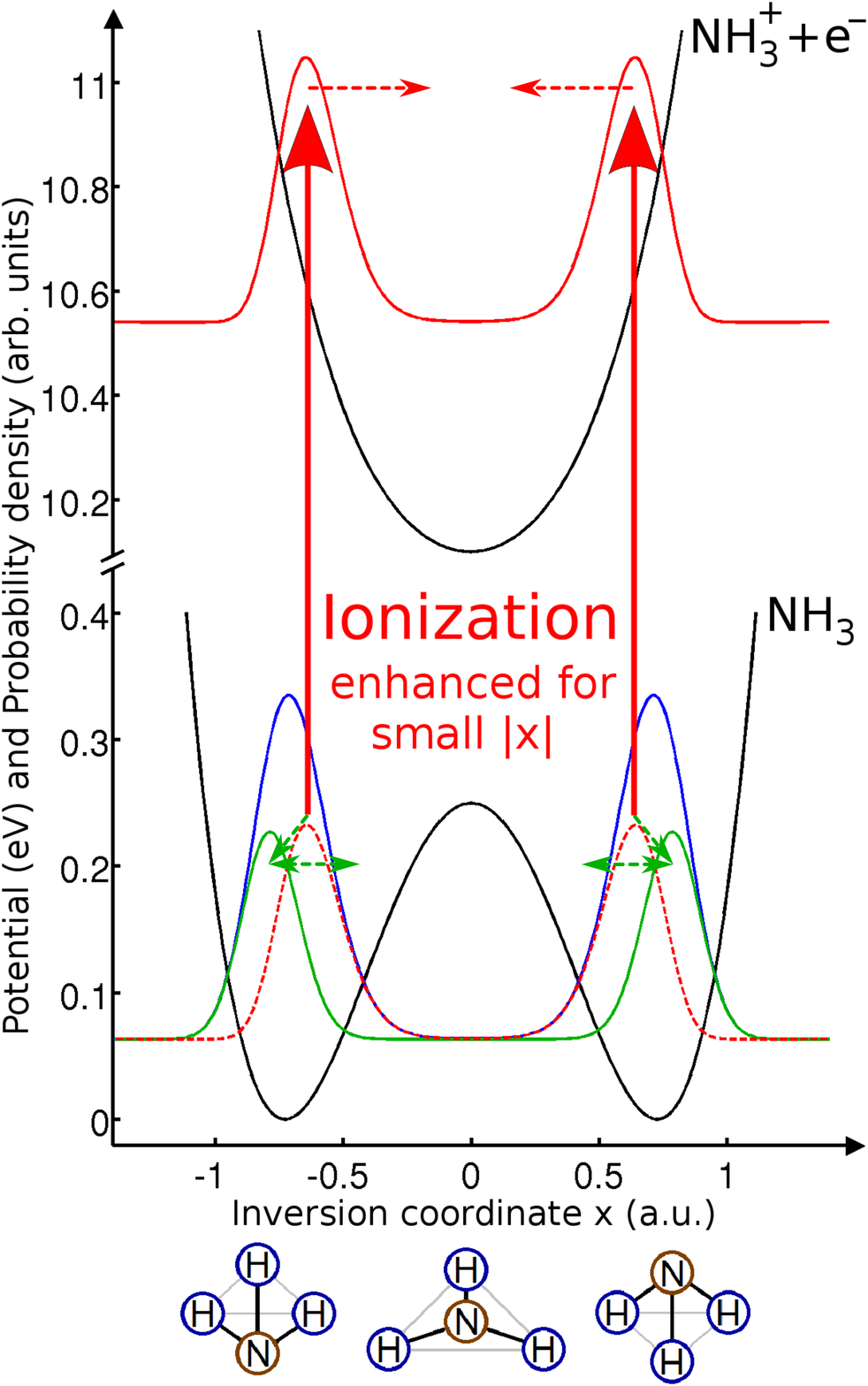}\\
\caption{\label{fig:WavepacketsSheme} (Color online) Potential energy
  curves of the electronic ground states of NH$_3$ and NH$_3^+$ (black
  lines) and probability density of the initial state (blue line). The
  full red line indicates the density generated in NH$_3^+$ due to
  ionization.  This part of the probability density is then missing in
  neutral NH$_3$ as indicated by the dashed red line, such that the
  density indicated by the green line is created.  }
\end{center}
\end{figure}

The ionization yields obtained from the solution of the TDSE as well
as from the ADK and FC-ADK rates are shown in
Fig.~\ref{fig:IonizationYieldsLong} for $n_{\mathrm{c}}=6-10$ cycle
pulses and wavelengths of $\lambda=800, 1300$, and $1800$ nm.  The
TDSE ionization yields are almost perfectly symmetric with respect to
the inversion coordinate $x$, i.e.\ the ionization yield is unchanged
whether the maximum of the electric field is oriented parallel or
antiparallel to the inversion axis (see Fig.~\ref{fig:dipole_effect}).
While ADK qualitatively predicts the correct geometry dependence of
the ionization yield, FC-ADK also agrees quantitatively very well with
the TDSE results over the whole intensity range, similar to
observations for the H$_2$ molecule \cite{sfm:foer14}. The largest
differences between ADK and FC-ADK are observable at low intensities
and $\lambda=800$ nm. The differences between ADK and FC-ADK decrease
with increasing wavelength and intensity due to the decreasing Keldysh
parameter. Furthermore, one can observe how the TDSE ionization yields
also become smoother for decreasing Keldysh parameters, 
i.\,e.\ multiphoton resonances get less important
when going deeper into the quasistatic regime.  All predictions show a
significant increase in the ionization yield with decreasing absolute
value of the inversion coordinate $|x|$.  This is expected due to the
decrease of the vertical binding energy $I_p(x)$ with decreasing $|x|$
which stems from the change of the equilibrium geometry from pyramidal to
planar upon ionization, see also Fig.~\ref{fig:WavepacketsSheme}. The
exponential dependence of the ionization rate on the binding energy,
Eq.~\eqref{eq:IonRateCor}, leads to this increase.  However, the TDSE
shows a less pronounced $|x|$ dependence than ADK and FC-ADK. Especially at
small $|x|$ the TDSE ionization yield behaves much flatter than in the
(FC-)ADK case.

\begin{figure*}
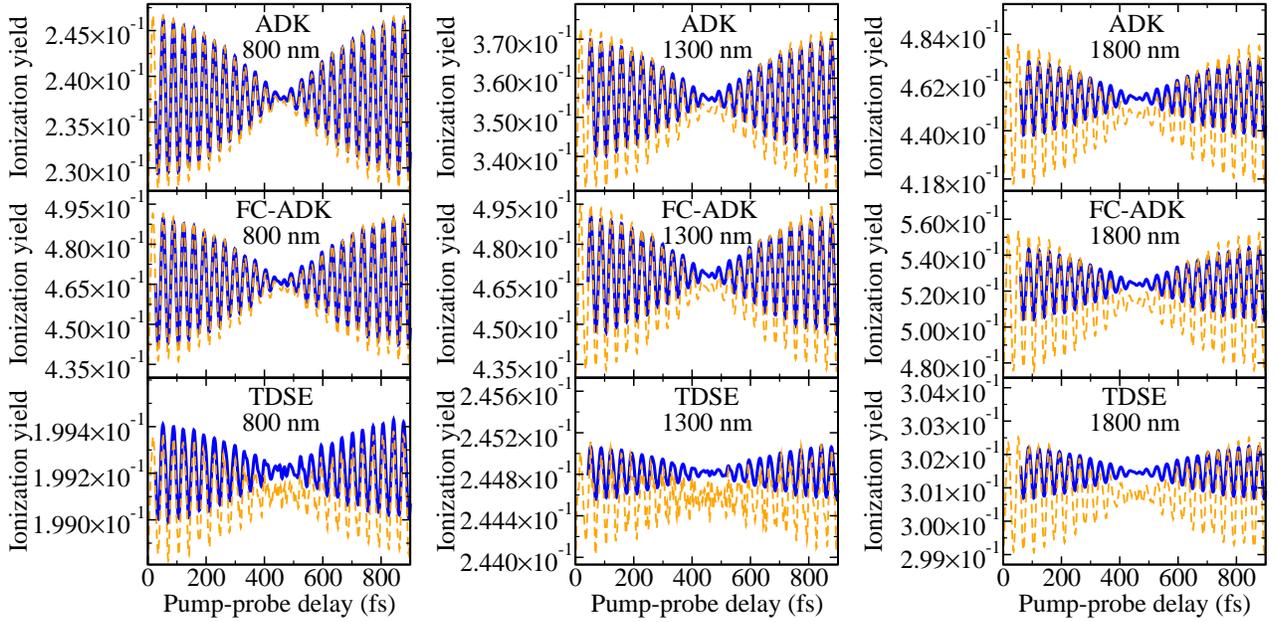

\begin{center}
     \includegraphics[width=0.3\textwidth]{figure06a.eps}\hspace*{0.2cm}
     \includegraphics[width=0.3\textwidth]{figure06b.eps}\hspace*{0.2cm}
     \includegraphics[width=0.3\textwidth]{figure06c.eps}
     \vspace*{-0.2cm}
     \caption{\label{fig:StdLochfrass} (Color online) Probe-pulse
       ionization yield as a function of the pump-probe delay.  A 10
       cycle $10^{14}$ W/cm${}^2$ pump pulse is followed by an
       identical 10 cycle $10^{14}$ W/cm${}^2$ probe pulse. Shown is
       the full time-dependent solution of
       Eq.~\eqref{eq:lochfrass-prop} (full blue lines) and the sudden
       ionization approximation, Eq.~\eqref{eq:sudden_approx} (dashed
       orange lines).}
\end{center}
\end{figure*}
\begin{figure}
\begin{center}
  \includegraphics[width=0.31\textwidth]{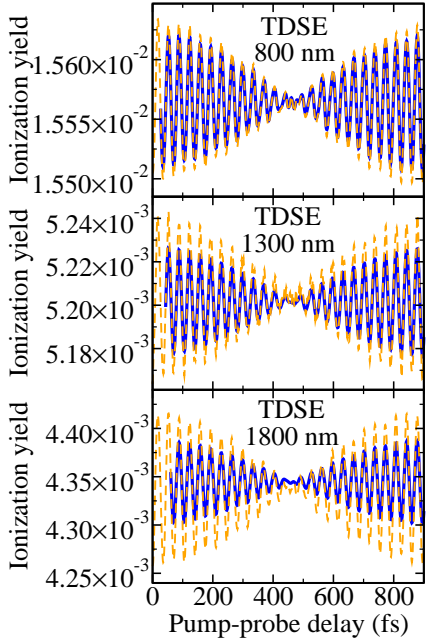}\vspace*{-0.2cm}
  \caption{\label{fig:EnhLochfrass} (Color online) As
    Fig.~\ref{fig:StdLochfrass}, but for a less intense 10-cycle
    $4\times 10^{13}$ W/cm${}^2$ probe pulse and showing the TDSE 
    ionization yields only.}
\end{center}
\end{figure}

We now consider Lochfra\ss\ similarly to previous studies for diatomic
molecules \cite{sfm:goll06, sfm:ergl06, sfm:fang08a, sfm:fang08b}.
Starting from ammonia in its thermal equilibrium state ($50\%
\left|0s\right\rangle$ and $50\% \left|0a\right\rangle$), the
geometry dependence of the ionization yield will create a wave packet
as sketched in Fig.~\ref{fig:WavepacketsSheme}. The wave packet
created in the NH$_3^+$ ion (Fig.~\ref{fig:WavepacketsSheme} top) is
shifted towards smaller values of $|x|$ due to the enhancement of the
ionization yield at smaller $|x|$.  This wave packet will then move
towards the potential minimum of NH$_3^+$, see \cite{sfm:foer13} for a
detailed study. Since the probability density remaining in neutral
NH$_3$ is mostly depleted at small values of $|x|$, a wave packet
shifted towards larger values of $|x|$ remains after ionization
(Fig.~\ref{fig:WavepacketsSheme} bottom). This wave packet will then
oscillate inside the minima of the double minimum potential. To probe
the wave packet in neutral NH$_3$, one can again exploit the geometry
dependence of the ionization yield. Whenever the wave packet is close
to the potential barrier around $x=0$ a.u., the total ionization
probability will be enhanced compared to positions further away from
the barrier. Thus, one can use a pump pulse to create the wave packet
and image it with a probe pulse. The created wave packet can then be
followed in real time by varying the pump-probe delay.
Fig.~\ref{fig:StdLochfrass} shows the ionization yield of the probe
pulse as a function of the pump-probe delay using two delayed,
identical $n_{\mathrm{c}}=10$ cycle pulses with a peak intensity of
$I=10^{14}$ W/cm${}^2$ and wavelengths $\lambda=800, 1300$, and $1800$
nm. In all cases a similar oscillation of the ionization yield is
observed.  This oscillation can easily be understood considering that
the symmetry with respect to $x$ is not broken at any time and that
the probability densities of $\left|0s\right\rangle$ and
$\left|0a\right\rangle$ are (almost) equal.  Thus, two independent
oscillations with initially same amplitude and phase are launched, one
for the symmetric and another one for the antisymmetric
part. Similarly to H$_2$~\cite{sfm:goll06}, mainly the first excited
state (but here of each symmetry) gets populated by the pump
pulse. The oscillations for each symmetry have a period of
\begin{align}
T_s&=\frac{2\pi\hbar}{E_{1s}-E_{0s}}\approx 35.80\ \mbox{fs}\quad  \mbox{and}\\
T_a&=\displaystyle \frac{2\pi\hbar}{E_{1a}-
  E_{0a}}\approx 34.47\ \mbox{fs}\quad .
\end{align}  
Due to this slight difference in their oscillation period (or
frequency), both contributions will cancel out after
\begin{align}
  \frac{T_{\mathrm{envelope}}}{2}\,=\,
  \frac{T_sT_a}{2\left(T_s-T_a\right)} \approx 462\ \mbox{fs} 
\end{align}
and revive after $T_{\mathrm{envelope}}$, as is seen in
Fig.~\ref{fig:StdLochfrass} for all cases.  Regarding the absolute
numbers, one can see that ADK and FC-ADK predict a significantly
larger contrast (oscillation height vs.\ average value of the
ionization yield) than the TDSE. This is due to the much smaller
$|x|$ dependence of the TDSE ionization yields for $I=10^{14}$
W/cm${}^2$, see Fig.~\ref{fig:IonizationYieldsLong}. Especially for
$800$\,nm, the contrast
$C=(Y_{\mathrm{max}}-Y_{\mathrm{min}})/Y_{\mathrm{average}}$ is
extremely small in the TDSE case, $C_{\mathrm{TDSE}} \approx 0.0023$,
compared to $C_{\mathrm{ADK}} \approx 0.072$ and $C_{\mathrm{FC-ADK}}
\approx 0.10$.  For $1800$\,nm, the contrast changes mainly for the
TDSE and is larger, $C_{\mathrm{TDSE}} \approx 0.0053$, but still more than an
order of magnitude smaller than $C_{\mathrm{FC-ADK}} \approx 0.078$. 
Even longer wavelengths could possibly improve the contrast further. 
The sudden ionization approximation for the nuclear motion,
Eq.~\eqref{eq:sudden_approx}, which is also shown in
Fig.~\ref{fig:StdLochfrass} agrees qualitatively very well (almost
quantitatively) to the time-dependent solution of
Eq.~\eqref{eq:lochfrass-prop}.  Noteworthy, for the TDSE results at
$800$ and $1300$ nm where the signal is very small, the sudden
ionization approximation shows a more "spikey" signal.

Fig.~\ref{fig:EnhLochfrass} considers a similar pump-probe scheme as
before, but with a lower probe-pulse intensity of
$I=4\times10^{13}$~W/cm${}^2$. Since the geometry dependence of the
TDSE ionization yield is larger for smaller intensities, one would
expect a higher contrast. Indeed, the contrast now reaches the
percentage regime, namely $C_{\mathrm{TDSE}} \approx~0.007\,(0.01,
0.02)$ for $800\, (1300, 1800)$\, nm. The contrast may be further
increased using even lower probe-pulse intensities. This would lead to
even lower total ionization yields which, however, may cause practical
problems due to experimental sensitivity limits. The optimal intensity
is thus a compromise between contrast and sensitivity and depends on 
the given experimental setup. 

%
%#############################################################################
\section{Imaging of a tunneling wave
  packet}\label{sec:tunneling-wavepacket}
%#############################################################################
%

Since the ionization yield gets larger whenever a wave packet moves
closer to the barrier of the double-well potential, one may think of
exploiting this behavior to directly observe a wave packet tunneling through the
barrier.  Considering a coherent superposition of the two lowest
vibrational eigenstates,
\begin{align}
  \left| \chi(t=0) \right\rangle\, =\,
  \frac{1}{\sqrt{2}}\left(\,\left| 0s \right\rangle\, +\, \left| 0a
    \right\rangle\,\right) \quad , \label{eq:tunneling-wp}
\end{align}
a wave packet tunneling from one side to the other is achieved.  This
wave packet can be realized considering that the ammonia maser
operates on the
$\left|0s\right\rangle\,\leftrightarrow\,\left|0a\right\rangle$
transition \cite{gen:gord54}.  Starting from a beam of
$\left|0a\right\rangle$ states within a maser set-up, the wave packet $\left| \chi(t=0)
\right\rangle$ is created after a quarter of the maser transition
period and may then propagate freely. Then, after
\begin{align}
  \tau\,=\,\frac{\pi\hbar}{E_{0a}-E_{0s}}\, \approx\, 20 \ \mbox{ps} \label{eq:total-time-tunneling}
\end{align}
this wave packet will have tunneled completely to the other side.  The
probability density $\rho(x,t)\,=\,\left|\chi(x,t)\right|^2$ of the
wave packet evolves according to
\begin{align}
  \rho(x,t)\,=\,\cos^2\left(\frac{\pi t}{2\tau
    }\right)\,\rho_{\mathrm{left}}(x)\,+\,\sin^2\left(\frac{\pi
      t}{2\tau
    }\right)\,\rho_{\mathrm{right}}(x)\label{eq:tunneling-density-TD}
\end{align}
where 
\begin{align}
  \rho_{\mathrm{left}}(x)\,=\,\rho(x,0)\,=\,\frac{1}{2}\,\big|\,\Psi_{0s}(x)\,+\,\Psi_{0a}(x)\,\big|^2
\end{align}
corresponds to the probability density localized on the left side of
the double-well potential and
\begin{align}
  \rho_{\mathrm{right}}(x)\,=\,\rho(x,\tau)\,=\,\frac{1}{2}\,\big|\,\Psi_{0s}(x)\,-\,\Psi_{0a}(x)\,\big|^2
\end{align}
to the density localized on the right side.
\begin{figure*}
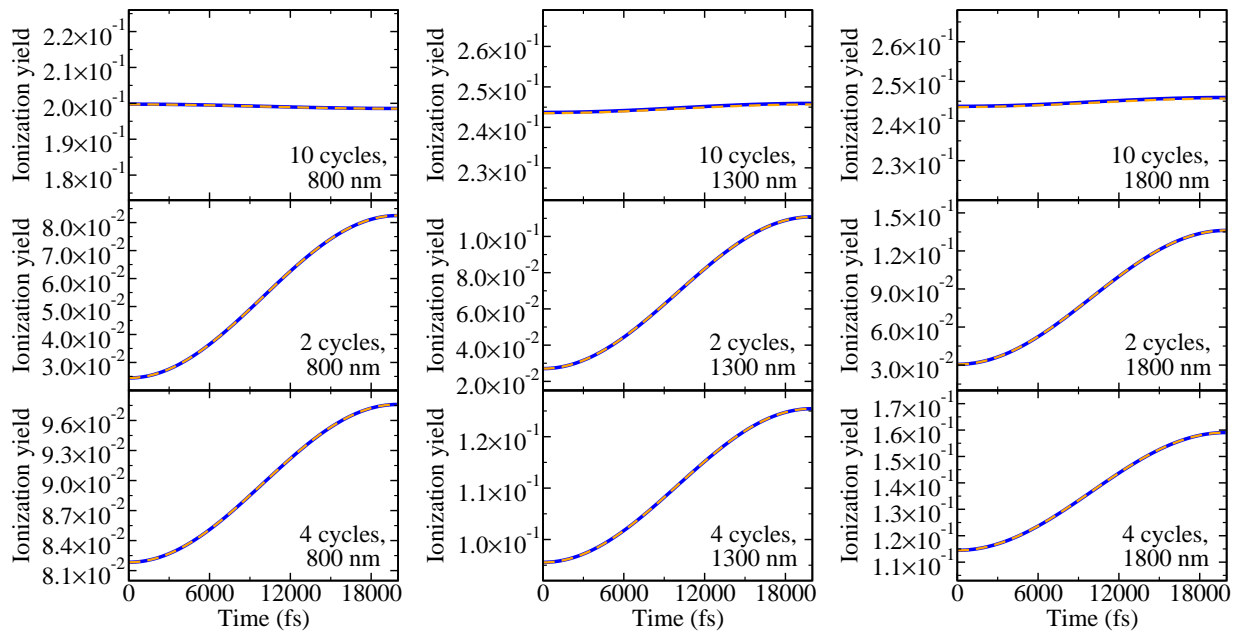

\begin{center}
         \includegraphics[width=0.29\textwidth]{figure08a.eps}\hspace*{0.2cm}
         \includegraphics[width=0.29\textwidth]{figure08b.eps}\hspace*{0.2cm}
         \includegraphics[width=0.29\textwidth]{figure08c.eps}
         \caption{\label{fig:tunnelingWP} (Color online) Ionization
           yield for $I=10^{14}$~W/cm${}^2$ pulses (number of cycles
           and wavelength given inside the graphs) probing the
           tunneling wave packet. Shown are the full time-dependent
           solutions of Eq.~\eqref{eq:lochfrass-prop} (full blue lines)
           and the results within the sudden ionization approximation,
           Eq.~\eqref{eq:sudden_approx} (dashed orange lines).}
\end{center}
\end{figure*}
\begin{figure}
\begin{center}
  \includegraphics[width=0.42\textwidth]{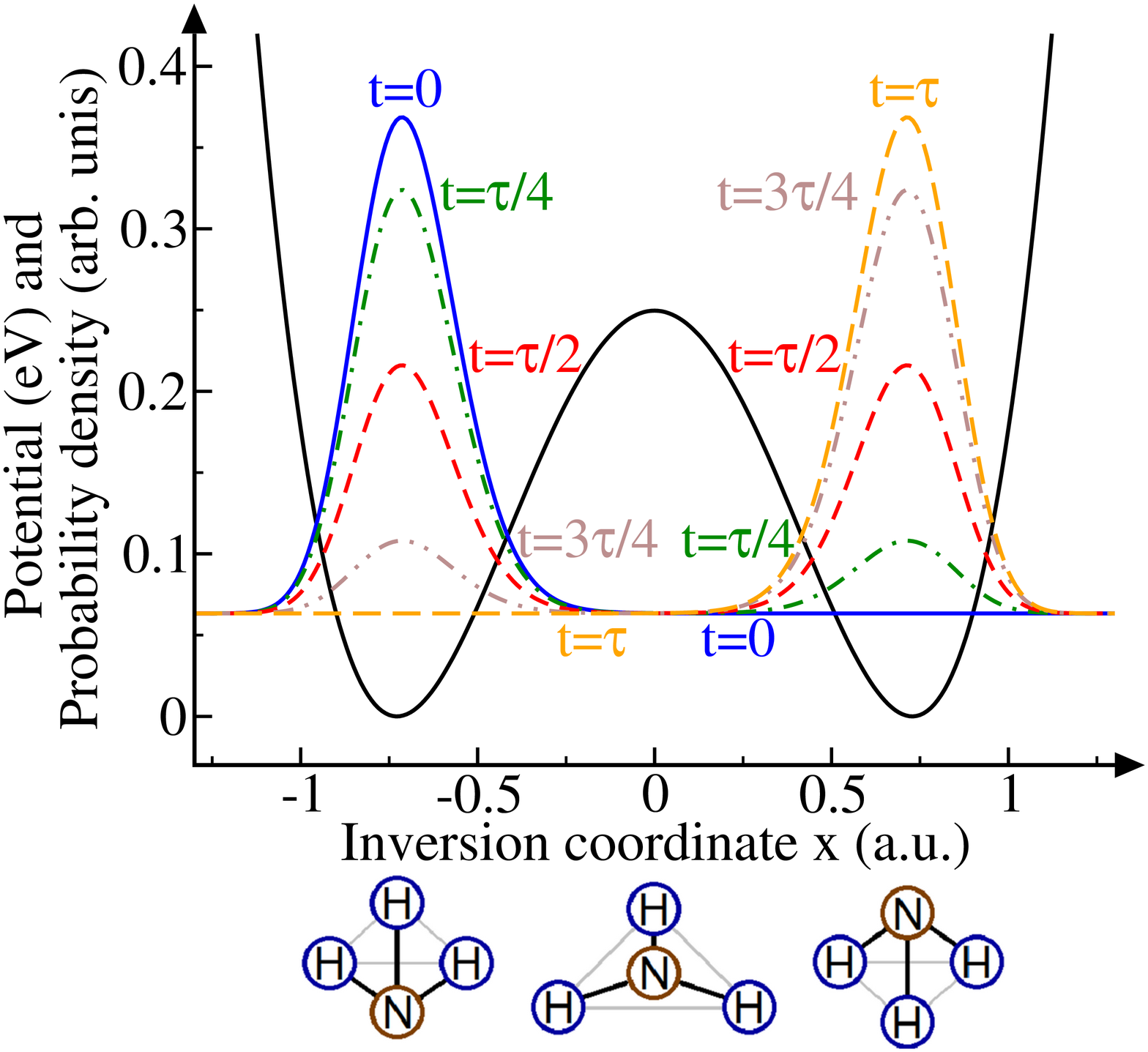}\\
  \caption{\label{fig:TunnelingWavePacket} (Color online) Snapshots of
    the time evolution of a tunneling wave packet (full blue line and
    dashed / chain lines) for $t=0, \tau/4, \tau/2, 3\tau/4,$ and
    $\tau$ in the ground state potential of NH$_3$ (full black line)
    where $\tau$ is the time required to tunnel from one well to the 
    other completely, Eq.~\eqref{eq:total-time-tunneling}.}
\end{center}
\end{figure}
If a single 10-cycle laser pulse is used to probe this wave packet,
the ionization yield remains, however, almost constant during the
tunneling process as is seen in the top row of
Fig.~\ref{fig:tunnelingWP}. Due to the symmetry of the tunneling
process, the ionization yield even stays perfectly constant for an
ionization yield $Y(x)$ which is symmetric in $x$ ($\leftrightarrow
-x$). This reflects the way how this wave packet tunnels according to
Eq.~\eqref{eq:tunneling-density-TD}.
Fig.~\ref{fig:TunnelingWavePacket} shows snapshots of the time
evolution of this wave packet. Instead of "moving through" the
barrier, the probability density rather disappears on one side of the
barrier and appears on the other. Thus, even what naively seems to be
the perfect detector for imaging the tunneling process, a symmetric
probe where the signal is larger when the probability density is
located further inside the tunneling barrier, yields only a constant
signal during the tunneling process.

%
%#############################################################################
\section{Breaking the symmetry: imaging with an asymmetric ionization
  yield}\label{sec:asymmetric-ionization}
%#############################################################################
%

\begin{figure*}
\begin{center}
  \includegraphics[width=0.95\textwidth]{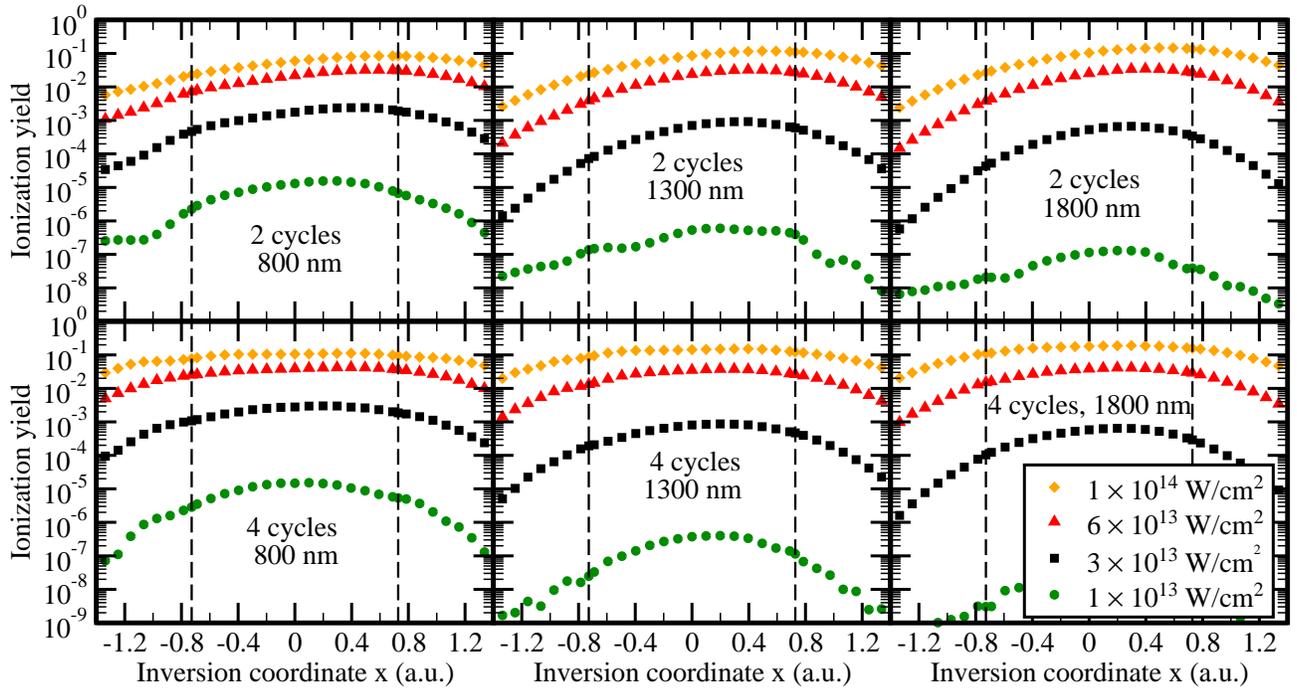}\\
  \caption{\label{fig:IonizationYieldsShort} (Color online) As
    Fig.~\ref{fig:IonizationYieldsLong}, but for shorter $n_c=2$ and
    $4$ cycle pulses and using TDSE ionization yields only.}
\end{center}
\end{figure*}

In order to follow the tunneling wave packet,
Eq.~\eqref{eq:tunneling-density-TD}, it is necessary to break the
symmetry of the ionization yield $Y(x)$ with respect to $x$.  This is
the case if for an oriented molecule the direction of the electric
field maximum with respect to the dipole moment (see
Fig.~\ref{fig:dipole_effect}) leads to significantly different
ionization yields for the parallel and antiparallel orientations as 
was demonstrated and discussed for the water molecule \cite{sfm:petr13a}.
Fig.~\ref{fig:IonizationYieldsShort} shows that for very short pulses
of only $n_c=2$ cycles, the TDSE ionization yields are indeed highly
asymmetric, for example $Y(x=0.73\ \mathrm{a.u.})/Y(x=-0.73\
\mathrm{a.u.})\,\approx\, 3.7\ (4.6,\ 5.1)$ for $800\ (1300,\ 1800)$
nm and $I=10^{14}$~W/cm${}^2$. For $n_c=4$ cycles, the ionization
yields are still notably asymmetric, e.g.\ $Y(x=0.73\
\mathrm{a.u.})/Y(x=-0.73\ \mathrm{a.u.})\,\approx\, 1.2\ (1.4,\ 1.4)$
for $800\ (1300,\ 1800)$ nm and $I=10^{14}$~W/cm${}^2$.  The asymmetry
is superimposed to the previously discussed enhancement of the
ionization yield for smaller $|x|$.  Note that this asymmetry is not
described within standard (FC-)ADK, Eq.~\eqref{eq:Y_ADK}, since the
binding energy $I_p(x)$ is perfectly symmetric with respect to $x$.
(A corresponding extension including dipole effects in the weak-field limit has, 
however, been suggested in \cite{sfm:tols11, sfm:tols14}.)

The second and third row of Fig.~\ref{fig:tunnelingWP} show the
time-dependence of the total ionization yield for the $n_c=2-4$ cycle
pulses from Fig.~\ref{fig:IonizationYieldsShort}.  As expected, the
ionization yield strongly increases with time, thus these pulses
indeed allow to follow the tunneling wave packet,
Eq.~\eqref{eq:tunneling-density-TD},
during the tunneling process. For these very short pulses, the sudden
ionization approximation agrees perfectly to the full solution of
Eq.~\eqref{eq:lochfrass-prop}.

\begin{figure}
\begin{center}
\includegraphics[width=0.38\textwidth]{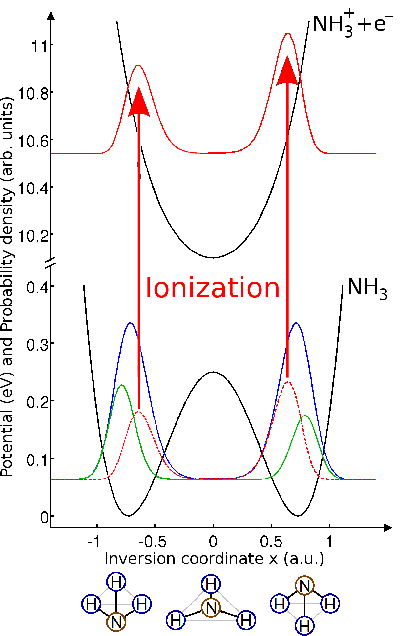}\\
\caption{\label{fig:WavepacketsShemeAsymmetric} (Color online) As
  Fig.~\ref{fig:WavepacketsSheme}, but for an asymmetric ionization
  yield.  }
\end{center}
\end{figure}

\begin{figure*}
\begin{center}
     \includegraphics[width=0.32\textwidth]{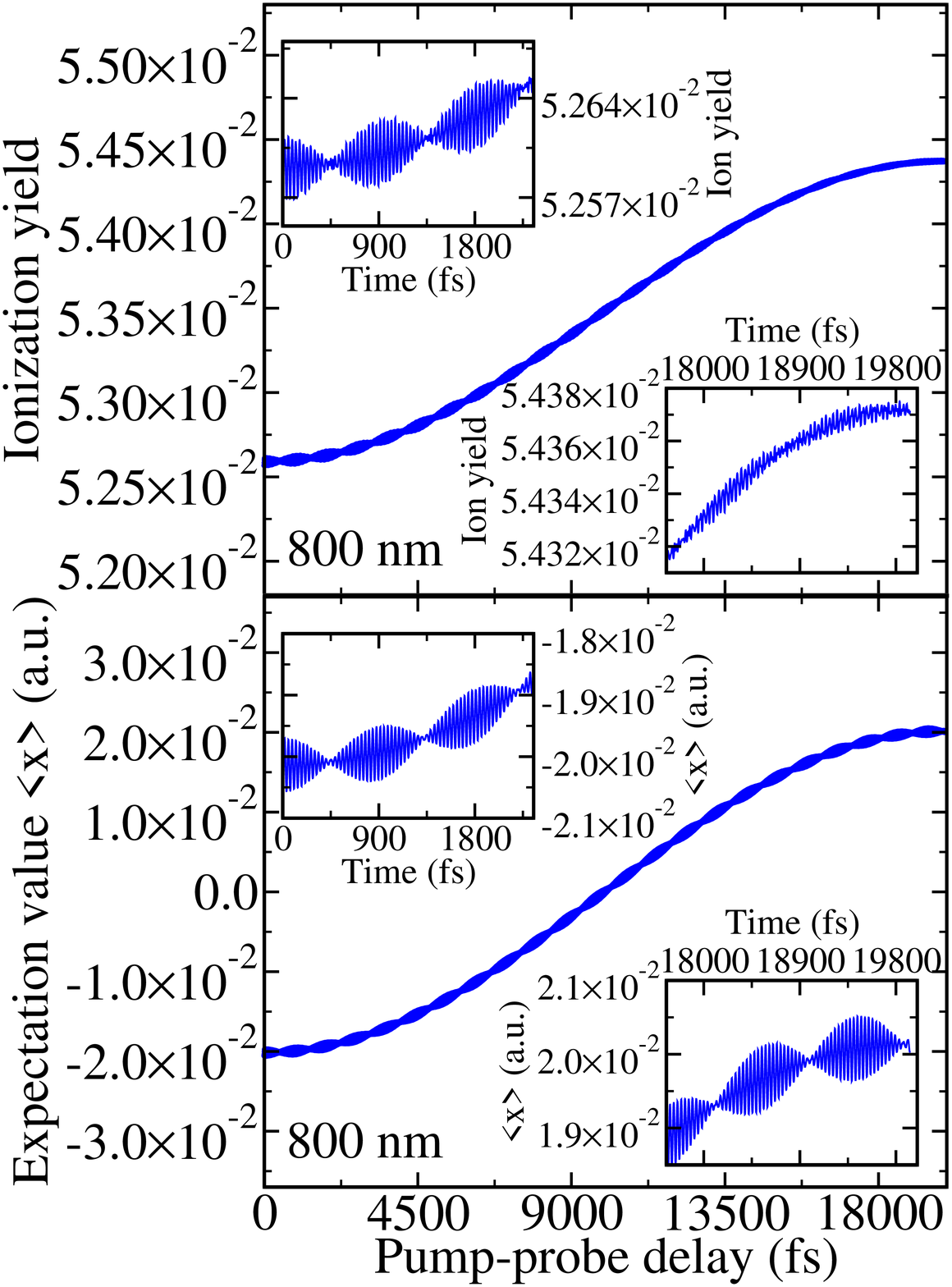}\hspace*{0.2cm}
     \includegraphics[width=0.32\textwidth]{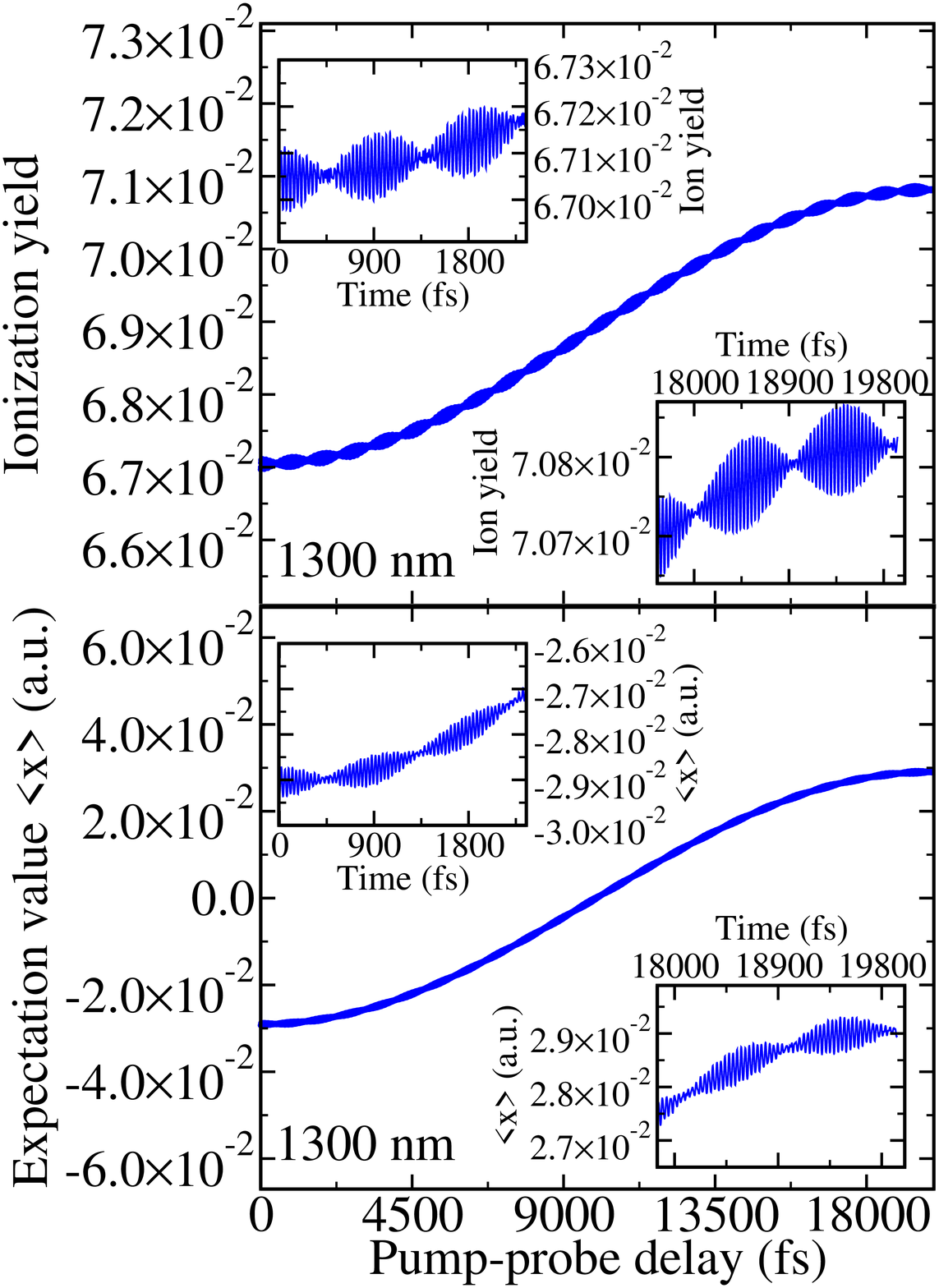}\hspace*{0.2cm}
     \includegraphics[width=0.32\textwidth]{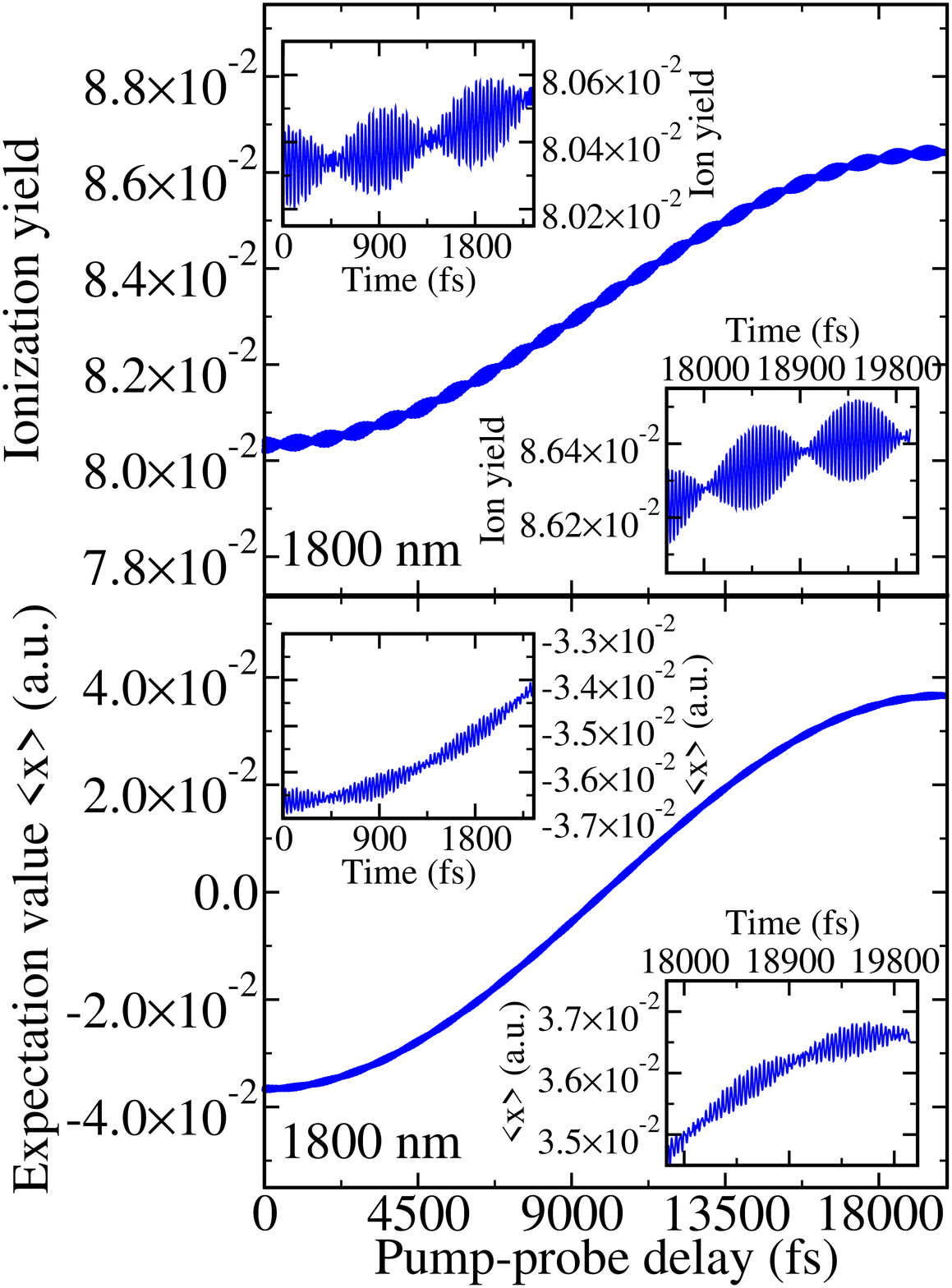}\\     
     \caption{\label{fig:SuperShortLochfrass} (Color online)
       Probe-pulse ionization yield as a function of the pump-probe
       delay for wavelengths $800, 1300,$ and $1800$\,nm.  A 2-cycle
       $10^{14}$ W/cm${}^2$ pump pulse is followed by an identical
       2-cycle $10^{14}$ W/cm${}^2$ probe pulse. Shown is the
       probe-pulse ionization yield obtained from the propagation of
       the density matrix (top) and the expectation value
       $\left\langle x \right\rangle$ of the created wave packet
       (bottom). The insets enlarge the signal at the beginning and
       the ending of the graph.}
\end{center}
\end{figure*}

An asymmetric wave packet in neutral NH$_3$ may also be created
starting from the $50\%\ \left|0s\right\rangle$, $50\%\
\left|0a\right\rangle$ thermal equilibrium state by ionizing with
an asymmetric ionization yield $Y(x)$, see
Fig.~\ref{fig:WavepacketsShemeAsymmetric}.  We now consider a
Lochfra\ss\ pump-probe scheme as previously discussed, but now using
two 2-cycle $10^{14}$ W/cm${}^2$ pulses. In this case, the density
matrix describing the ensemble is propagated within the sudden
ionization approximation since an independent propagation of
$\left|0s\right\rangle$ and $\left|0a\right\rangle$ is not possible
for an asymmetric ionization yield. However, as previously noted, the
sudden ionization approximation agrees perfectly to the solution of
the TDSE for these very short
pulses. Fig.~\ref{fig:SuperShortLochfrass} shows the obtained
probe-pulse ionization yields. The behavior of the probe-pulse
ionization yield follows the expectation value $\left\langle x
\right\rangle$ of the wave packet.  The created asymmetry leads to a
behavior similar to the tunneling wave packet in
Fig.~\ref{fig:tunnelingWP}, but with a significantly smaller contrast
since the asymmetry created by the pump pulse is smaller.
Superimposed to the signal from tunneling is an oscillation similar to
the signal for 10-cycle pulses, Fig.~\ref{fig:StdLochfrass}, due to
the $|x|$ dependence of the ionization yield,
Fig.~\ref{fig:IonizationYieldsShort}. The contrast of this
sub-structure depends on the $|x|$ dependence of the ionization yield
on either side.  For example, the very flat behavior of the ionization
yield for $n_c=2$ cycles, $\lambda=800$ nm and $I=10^{14}$ W/cm${}^2$,
Fig.\ref{fig:IonizationYieldsShort}, leads to a rather small
sub-oscillation of the ionization yield at the end of the tunneling
process compared to its beginning, see
Fig.~\ref{fig:SuperShortLochfrass}.

Note that an asymmetric ionization behavior may alternatively also 
be observed for longer pulses as an asymmetry in the emission direction
of the electrons and thus if the photoelectron 
spectra are recorded, or may be created by $\omega/2\omega$ pulses. 
Therefore, a number of possible experimental realizations of directly 
observing the tunneling process in ammonia exists.

%
%#############################################################################
\section{Conclusions}
%#############################################################################
%

The geometry-dependent ionization behavior of ammonia has been studied
by solving the TDSE within the single-determinant approximation.  For
10-cycle pulses ADK qualitatively predicts the correct ionization
yield, while FC-ADK agrees to the TDSE almost quantitatively over the
whole intensity range.  However, both ADK and FC-ADK significantly
overestimate the increase of the ionization yield with decreasing
inversion coordinate $|x|$.  Thus, while similar vibrational dynamics
through Lochfra\ss\ is observable using TDSE or (FC-)ADK ionization
yields, the contrast of the oscillation of the ionization yield is
much smaller for the TDSE ionization yield compared to (FC-)ADK.  The
contrast can nevertheless be improved using $1800$ nm instead of $800$
nm or a weaker probe pulse.

Furthermore, imaging of a tunneling wave packet has been discussed.
While a detector leading to an increased ionization yield for wave
packets close to the tunneling barrier may seem perfect to follow the
tunneling process in real time, any geometry-dependent ionization
yield symmetric in $x$ leads to a perfectly constant signal.  This can
be understood looking at how the probability density evolves during
the tunneling process.  It was shown that very short 2-4 cycle pulses
break the symmetry of the geometry-dependent ionization yield with
respect to $x$, allowing to follow the tunneling wave packet in real
time.  Finally, Lochfra\ss\ using two 2-cycle pulses shows both the
behavior of the tunneling wave packet and the faster beating oscillations
previously discussed for the longer 10-cycle pulses.

Clearly, an experimental measurement of the predictions made here
would pave the way for future real-time movies of quantum-mechanical
nuclear motion in more than diatomic neutral molecules.  The results
show that FC-ADK may give a good estimate for laser parameters in
order to observe Lochfra\ss\ also for other, especially more complex
molecules which undergo an equilibrium geometry change upon
ionization, though the contrast may be overestimated. For oriented
molecules, asymmetries in the ionization behavior allow for even more
applications such as the real-time movie of the tunneling process
discussed here.  One may even be able to follow proton or heavy-atom 
tunneling in biologically relevant molecules 
\cite{gen:loew63, gen:masg06, gen:gonz10, gen:alle09, gen:arnd09}, 
provided the ionization yield of the orbital dominating the strong-field 
ionization process (usually the HOMO) changes sufficiently as a function 
of the tunneling coordinate.

\begin{acknowledgments}
  We gratefully acknowledge many helpful discussions with 
  and advice by P.~Decleva as well as 
  grants of computer time from CINECA.  J.~F. gratefully acknowledges
  a PhD scholarship from the {\it German National Academic Foundation
    (Studienstiftung des deutschen Volkes)}. We gratefully acknowledge
  financial support from the {\it Deutsche Forschungsgemeinschaft
    (DFG)} within Priority Programme 1840 QUTIF, the {\it EU Initial
    Training Network (ITN) CORINF}, and the {\it European COST Action
    CM1204 (XLIC)}.
\end{acknowledgments}

\bibliographystyle{apsrev}
%\bibliography{journals,sfm,sfa,dia,gen}

\end{document}